\documentclass[aps,prb,twocolumn,showpacs,eqsecnum]{revtex4-1}
\usepackage[latin1]{inputenc}
\usepackage[T1]{fontenc}
\usepackage{graphicx}
\usepackage{amsmath}
\usepackage{amssymb}
\usepackage{bbm}
\usepackage{dsfont}
\usepackage{feynmp}
\usepackage{nicefrac}

\newcommand{\bq}{\mathbf{q}}

\newcommand{\bp}{\mathbf{p}}

\newcommand{\bpp}{\mathbf{p}^\prime}
\newcommand{\ii}{\mathrm{i}}

\begin{document}

\title{Possibility of superconductivity due to electron-phonon
interaction in graphene}
\author{Matthias Einenkel and Konstantin B. Efetov}
\affiliation{Institut f\"ur Theoretische Physik III, Ruhr-Universit\"at Bochum, 44780
Bochum, Germany}
\date{\today }

\begin{abstract}
We discuss the possibility of superconductivity in graphene taking into
account both electron-phonon and electron-electron Coulomb interactions. The
analysis is carried out assuming that the Fermi energy is far away from the
Dirac points, such that the density of the particles (electrons or holes) is
high. We derive proper Eliashberg equations that allow us to estimate the
critical superconducting temperature. The most favorable is pairing of
electrons belonging to different valleys in the spectrum. By using values of
electron-phonon coupling estimated in other publications we obtain the
critical temperature $T_{c}$ as a function of the electron (hole) density.
This temperature can reach the order of $10$ \textrm{K} at the Fermi energy
of order $1-2$ eV. We show that the dependence of the intervalley pairing on
the impurity concentration should be weak.
\end{abstract}

\maketitle

\section{Introduction}

Since its experimental discovery in 2004~\cite{geim2004} and first
observations of unusual properties,~\cite{novoselov2005,kim2005} graphene
has gained a lot of experimental and theoretical attraction. In the last
decade, thousands of articles devoted to the study of graphene have
appeared. Possessing novel electro-mechanical properties, graphene is a
promising material for electronic devices. The linear Dirac-type electron
spectrum makes graphene very interesting from the theoretical point of view
and many interesting effects have been predicted and observed.~\cite%
{castro}

However, still there are open questions on fundamental electronic properties
of graphene, and one of the most interesting ones concerns a possibility of
superconductivity. Graphene is a good conductor unless the Fermi energy is
too close to the Dirac points and the electron-phonon coupling in graphene
is not very weak. Therefore, although the superconductivity in graphene has
not been observed, it is not clear why one should discard the possibility of
this phenomenon.

Superconductivity can be induced in graphene by superconducting contacts due
to the proximity effect,~\cite{heersche} but can it be obtained in a
\textquotedblleft natural way"? What should one do in order to obtain a
considerable value of the superconducting critical temperature $T_{c}$? What
type of the superconductivity and what structure of the order parameter
could one expect?

In the last years, there have been various attempts to answer these
questions. Due to the special type of the spectrum, the main attention has
been devoted to investigating the possibility of unusual types of the
superconducting pairing. Superconducting pairing mediated by conventional
electron-phonon or electron-plasmon coupling~was considered in Ref.~[%
\onlinecite{uchoa}] with a conclusion that, in addition to the conventional $%
s $-wave pairing, an exotic $p+ip$ state is possible. Superconducting
properties of Dirac electrons in graphene were considered in Ref.~[%
\onlinecite{kopnin}] within the conventional BCS approach. In these
publications, the main emphasis was put on the study of properties of unusual
superconducting pairing for a small electron density.

It is clear that one can expect very interesting new properties of the
superconductivity in the vicinity of the Dirac points. However, this region
is least favorable for the existence of superconductivity due to the very
low density of states, and one should tune the Fermi energy away from the
Dirac points in order to have a hope to obtain superconductivity.

By doping graphene by various combinations of K and Ca, the authors of Ref.~[%
\onlinecite{chesney}] were able to shift the Fermi energy far away the Dirac
points and even to put it in the vicinity of the van Hove singularity~(VHS).
Another experimental method based on the use of electrolytic gates \cite%
{efetov} allowed the authors to tune continuously the electron density up to
values $n=4\times 10^{14}\mathrm{cm}^{-2}$, which is apparently not very far
away from the VHS. These experimental works have demonstrated that one can
achieve a ultrahigh electron density and this makes observation of
superconductivity in graphene considerably more realistic. At the same time,
transport measurements were not carried out in Ref.~[\onlinecite{chesney}]
and the superconductivity has not been seen in Ref.~[\onlinecite{efetov}]
for temperatures $T\gtrsim 1\ \mathrm{K}$.

Although the superconductivity has not been observed yet, theoretical
considerations \cite{gonzalez,chubukov,thomale} predict superconductivity at the VHS
even for a repulsive electron-electron interaction. In this case, the
superconductivity is expected to have an unconventional symmetry of the
order parameter. No doubt, an experimental observation of the
superconductivity at the VHS would be of a great interest, but one should be
able to tune the Fermi energy exactly to the singularity. Disorder may also
play a destructive role in formation of such a superconductivity.

Therefore, it would still be important to investigate theoretically the
possibility of a superconducting pairing due to the conventional
electron-phonon pairing far away from the Dirac point, but, at the same time,
not in the vicinity of the VHS. Such a study implies using conventional
schemes of computing the superconducting transition temperature. Then, one
should check the stability of the pairing against disorder in the system,
clarify the dependence of the transition temperature $T_{c}$, etc.

In several publications, models with an electron-phonon interaction have been
considered. The authors of Ref.~[\onlinecite{lozovik}] discussed the valley
structure of the order parameter using an electron-phonon model without the
electron-electron Coulomb interaction. They argued that there might be a
superconducting instability in highly doped graphene, while the valley
structure depends on the parameters of the electron-phonon interaction.
Superconductivity in hydrogenized graphene (graphane) has been considered as
well.~\cite{graphane} In this system, a model based on electron-phonon
interaction was used and a transition temperature of $90\ \mbox{K}$ in
p-doped graphane was predicted. At the same time, it is clear that taking
into account the Coulomb interaction is very important because it can in
principle be even stronger than the electron-phonon interaction. Moreover,
considering the latter in the weak coupling limit can also lead to incorrect
predictions.

In this paper, we use a generic model including both the electron-phonon and
electron-electron Coulomb interactions. Using Eliashberg-type equations,\cite%
{eliashberg,morel,mcmillan} we derive an expression for the transition
temperature determined by the electron and phonon interactions in graphene.
By using experimental values and results of numerical calculations for the
interactions obtained for the normal state, we estimate the transition
temperature $T_{c}$ and conclude that the superconductivity is possible with
$T_{c}$ of the order of several Kelvin. Most favorable is a singlet pairing
between different valleys. We show that such a pairing is not very sensitive
to disorder.

The paper is organized as follows. In Sec.~$\mathrm{II}$, we formulate the
Hamiltonian of quasiparticles in doped graphene describing the interaction
of quasiparticles with phonons and the Coulomb interaction. In Sec.~$\mathrm{%
III}$, we consider the electron pairing by deriving and solving the
Eliashberg equations. The effects of impurity scattering are considered in
Sec.~$\mathrm{IV}$. In Sec.~$\mathrm{V}$, estimates are made, and the
concluding Sec.~$\mathrm{VI}$ is devoted to discussions.

\section{Model Hamiltonian}

In this section, we introduce the Hamiltonian of interacting quasiparticles
in a single layer of graphene. For the undoped system, the Fermi surface
consists  of two nonequivalent points $\mathbf{K}$ and $\mathbf{K}%
^{\prime }=-\mathbf{K}$, called the Dirac points.~\cite{castro} The
quasiparticles around these points have a linear spectrum. This
approximation remains valid up to quite high energies. We consider in this
work doping levels corresponding to the Fermi energy $\varepsilon _{F}$ of
the order of $1\mathrm{eV}$. In other words, we consider the case when the
Fermi energy is sufficiently far away from both the Dirac point and the VHS.
In this region of parameters the spectrum consists of two well resolved
valleys and each valley is a cone. As the Fermi energy is far away from the
Dirac points one does not need to account for effects specific for the Dirac
equation.

The two valleys are numerated by a variable called \emph{isospin}. Due to
the electron-phonon and electron-electron interactions there are matrix
elements of the Hamiltonian mixing these two valleys. As we do not
investigate properties of the system near the Dirac point we do not use the
Dirac-type representation of the Schrodinger equation.

The Hamiltonian describing the electron-phonon system can be written in a
general form%
\begin{equation}
\hat{H}=\hat{H}_{0}+\hat{H}_{e,ph}+\hat{H}_{e,e},  \label{a01}
\end{equation}%
where
\begin{equation}
\hat{H}_{0}=\sum_{\mathbf{p,}\sigma }\left( \varepsilon \left( \mathbf{p}%
\right) -\mu \right) c_{\mathbf{p},\sigma }^{\dag }c_{\mathbf{p},\sigma }
\label{a0}
\end{equation}%
is the operator of the kinetic energy, $\varepsilon \left( \mathbf{p}\right)
$ is the spectrum of the non-interacting electrons, $\mu $ is the chemical
potential (Fermi energy at low temperatures), $\sigma $ is the spin index, $%
c_{\mathbf{p},\sigma }\left( c_{\mathbf{p},\sigma }^{\dag }\right) $ is the
electron annihilation (creation) operator for the electron with momentum $%
\mathbf{p}$ and spin $\sigma $, $\hat{H}_{e,ph}$ stands for the
electron-phonon interaction, and $\hat{H}_{e,e}$-for the electron-electron
one.

\subsection{Electron-phonon interaction}

The Hamiltonian $\hat{H}_{e,ph}$ describing the interaction between the
electrons and phonons can be written in a general form as~\cite{agd,grimvall}

\begin{equation}
\hat{H}_{e,ph}=\sum_{\mathbf{p,q},j,\sigma }g_{\mathbf{p,q},j}\ \Phi_{%
\mathbf{q},j}c_{\mathbf{p+q},\sigma }^{\dag }c_{\mathbf{p},\sigma },
\label{a1}
\end{equation}%
where $g_{\mathbf{p},\mathbf{q},j}$ is the electron-phonon coupling function
and $\Phi_{\mathbf{q},j}$ is the phonon field operator for the mode~$j$. As
usual, the most important contributions to the thermodynamics are expected
from the vicinity of the Fermi surface that consists in the case involved of
two circles.

In order to distinguish explicitly between the valleys we write $c_{\mathbf{K%
}+\mathbf{p},\sigma }\equiv a_{1,\mathbf{p},\sigma }$ and $c_{\mathbf{K}%
^{\prime }+\mathbf{p},\sigma }\equiv a_{2,\mathbf{p},\sigma }$ for
quasiparticles in the vicinity of the Fermi circles of the two valleys $1$
and $2$, where $a_{\alpha ,\mathbf{p,}\sigma },$ $\alpha =1,2$ are fermionic
annihilation operators with the momentum $\mathbf{p}$ measured from the
Dirac point of the $\alpha $-valley.

Using these notations we write the electron-phonon interaction $\hat{H}%
_{e,ph}$, Eq. (\ref{a1}), in the form
\begin{equation}
\hat{H}_{e,ph}=\sum_{\alpha ,\beta =1}^{2}\sum_{\mathbf{p,q,}j,\sigma }g_{%
\mathbf{p,q},j}^{\alpha \beta }\ \Phi_{\mathbf{q},j}^{\alpha \beta
}a_{\alpha ,\mathbf{p+q},\sigma }^{\dag }a_{\beta ,\mathbf{p},\sigma }.
\label{a3}
\end{equation}%
In Eq.~(\ref{a3}) the coupling constants $g^{\alpha \beta }$ and the phonon
field operators are related to those in Eq. (\ref{a1}) as
\begin{eqnarray}
g_{\mathbf{p,q},j}^{11}\ &=&g_{\mathbf{p-K,q},j},\quad g_{\mathbf{p,q}%
,j}^{22}\ =g_{\mathbf{p-K}^{\prime }\mathbf{,q},j},\quad  \label{a3a} \\
g_{\mathbf{p,q,}j}^{12} &=&g_{\mathbf{p-K}^{\prime }\mathbf{,q-Q},j},\quad
g_{\mathbf{p,q},j}^{21}=g_{\mathbf{p-K},\mathbf{q+Q},j},  \notag \\
\Phi_{\mathbf{q},j}^{11} &=&\Phi_{\mathbf{q},j},\quad \Phi_{\mathbf{q,}%
j}^{22}=\Phi_{\mathbf{q,}j},\quad  \notag \\
\Phi_{\mathbf{q},j}^{12} &=&\Phi_{\mathbf{q-Q},j},\quad \Phi_{\mathbf{q}%
,j}^{21}=\Phi_{\mathbf{q+Q},j} ,  \notag
\end{eqnarray}%
where $\mathbf{Q}=\mathbf{K}^{\prime }-\mathbf{K}$ is the vector connecting
the Dirac points. The phonon fields $\Phi$ are real in the coordinate
representation and therefore we obtain the following relations%
\begin{eqnarray}
\Phi_{-\mathbf{q},j}^{11} &=&\left( \Phi_{\mathbf{q},j}^{11}\right) ^{\ast
},\quad \Phi_{-\mathbf{q,}j}^{22}=\left( \Phi_{\mathbf{q,}j}^{22}\right)
^{\ast } ,  \label{a3b} \\
\Phi_{-\mathbf{q},j}^{12} &=&\left( \Phi_{\mathbf{q},j}^{21}\right) ^{\ast
},\quad \Phi_{-\mathbf{q},j}^{21}=\left( \Phi_{\mathbf{q},j}^{12}\right)
^{\ast }.  \notag
\end{eqnarray}

In Eq.~(\ref{a3}) the sum is taken over such $\mathbf{p}$ and $\mathbf{q}$
that both $\mathbf{p}$ and $\mathbf{p}+\mathbf{q}$ are in the vicinity of
the Fermi circle. The terms with $\alpha =\beta $ in the Hamiltonian $\hat{H}%
_{e,ph}$, Eq.~(\ref{a3}) describes the \emph{intravalley scattering} of
electrons by the phonons, while the terms with $\alpha \neq \beta $ stand
for the \emph{intervalley scattering}$.$ Formally, we can speak of an
isospin dependence of the electron-phonon interaction.

The bare Hamiltonian $\hat{H}_{0}$, Eq.~(\ref{a0}), takes in these notations
the form
\begin{equation}
\hat{H}_{0}=\sum_{\alpha =1}^{2}\sum_{\mathbf{p,}\sigma }\xi _{\mathbf{p}}{%
\alpha }_{\alpha ,\mathbf{p,}\sigma }^{\dag }{\alpha }_{\alpha ,\mathbf{p,}%
\sigma },  \label{a4}
\end{equation}%
where the energy $\xi _{\mathbf{p}}$ equals
\begin{equation}
\xi _{\mathbf{p}}=v_{0}\left\vert \mathbf{p}\right\vert -\mu ,  \label{a5}
\end{equation}%
and $v_{0}\approx 10^{8}~\mathrm{cm}\ \mathrm{s}^{-1}$ is the Fermi velocity.

Equations~(\ref{a3})-(\ref{a4}) specify how the presence of the two valleys can be
written in terms of the isospin. The coupling constants $g^{\alpha \beta }$
are different for the intravalley and intervalley scattering (equal or unequal $%
\alpha ,\,\beta $, respectively). In graphene, two two-phonon peaks are seen
in the Raman spectrum (see, e.g., Refs.~\onlinecite{ferrari,yan}): the $D^{\ast }$ peak
near $2\omega _{A_{1}}=2650 \ \mathrm{cm}^{-1}$ corresponding to the scalar $%
A_{1}$ optical phonons and the $G^{\ast }$ peak near $2\omega _{E_{2}}=3250
\ \mathrm{cm}^{-1}$ corresponding to twofold-degenerate pseudovector $E_{2} $
phonon~mode. The $E_{2}$ mode is responsible for intravalley scattering,
while the scalar $A_{1}$~optical mode leads to the intervalley scattering.~%
\cite{basko}

\subsection{Coulomb interaction}

The operator of the electron-electron interaction $\hat{H}_{e,e}$ in Eq.~(%
\ref{a01}) can be written in the standard form

\begin{equation}
\hat{H}_{e,e}=\frac{1}{2}\sum_{\mathbf{p},\mathbf{p}^{\prime },\mathbf{q}}\
V_{\mathbf{q}}\ c_{\mathbf{p+q}}^{\dag }c_{\mathbf{p}^{\prime }-\mathbf{q}%
}^{\dag }c_{\mathbf{p}^{\prime }}c_{\mathbf{p}},  \label{a6}
\end{equation}%
where $V_{\mathbf{q}}$ is the matrix element of the Coulomb potential in two
dimensions. We assume here that corrections to the bare Coulomb potential
have already been calculated and use therefore for $V_{\mathbf{q}}$ the
static screened Coulomb interaction. Due to a specific form of the wave
functions of graphene leading to a suppression of the backscattering, the
function $V_{\mathbf{q}}$ differs from the conventional Fourier transform of
the Coulomb interaction and can be written in the form%
\begin{equation}
V_{\mathbf{q}}=\frac{2\pi e^{2}}{\kappa \left\vert \mathbf{q}\right\vert
\epsilon (\mathbf{q})}\gamma (\mathbf{q})  \label{a7}
\end{equation}%
where $\kappa $ is the dielectric constant of the substrate, $\epsilon (%
\mathbf{q})=1+\frac{2\pi e^2}{\kappa \left\vert \mathbf{q} \right\vert} \Pi
\left( \mathbf{q}\right) $ is the static dielectric permeability of the
electron gas and
\begin{equation}
\gamma (\mathbf{q})=\frac{1}{2}\left( 1+\cos \phi _{\mathbf{q}}\right)
\label{a7b}
\end{equation}%
with $\phi _{\mathbf{q}}=\arctan \frac{q_{y}}{q_{x}}$ is the form~factor
accounting for the absence of backscattering in graphene. The static
polarizability $\Pi \left( \mathbf{q}\right) $ is given by \cite{Hwang}%
\begin{align}
\frac{\Pi (\mathbf{q})}{4\nu }=\qquad \qquad \qquad \qquad \qquad\qquad
\qquad  \notag \\
\left\{
\begin{array}{ccc}
1 & \mathrm{for} & q\leq 2p_{F} \\
1-\frac{1}{2}\sqrt{1-\frac{4p_{F}}{q}}-\frac{q}{4p_{F}}\sin ^{-1}\left(
\frac{2p_{F}}{q}\right) +\frac{\pi q}{8p_{F}} & \mathrm{for} & q > 2p_{F}%
\end{array}%
\right.  \label{a7a}
\end{align}%
where
\begin{equation}
\nu \left( \mu \right) =\frac{\mu }{2\pi v_{0}^{2}}  \label{DOS}
\end{equation}%
is the density of states per one spin direction and one valley and $p_{F}$
is the Fermi momentum. The factor $4$ in Eq.~(\ref{a7a}) accounts for the
number of the spin and isospin directions.

Again, we can use the isospin formulation to bring the operator $\hat{H}%
_{e,e}$ to a more convenient form
\begin{equation}
\hat{H}_{e,e}=\frac{1}{2}\sum_{\alpha ,\beta }\sum_{\mathbf{p},\mathbf{p}%
^{\prime },\mathbf{q}}V_{\mathbf{q}}^{\alpha \beta }a_{\alpha ,\mathbf{p+q}%
}^{\dag }a_{\beta ,\mathbf{p}^{\prime }-\mathbf{q}}^{\dag }a_{\alpha ,%
\mathbf{p}^{\prime }}a_{\beta ,\mathbf{p}},  \label{a8}
\end{equation}%
where%
\begin{eqnarray}
V_{\mathbf{q}}^{11} &=&V_{\mathbf{q}}^{22}=V_{\mathbf{q}},  \label{a9} \\
V_{\mathbf{q}}^{12} &=&V_{\mathbf{q-Q}},\quad V_{\mathbf{q}}^{21}=V_{\mathbf{%
q+Q}},  \notag
\end{eqnarray}%
and the sum in Eq.~(\ref{a8}) includes such momenta $\mathbf{p},\mathbf{p}%
^{\prime }$, and $\mathbf{q}$ that both $\mathbf{p+q}$ and $\mathbf{p}%
^{\prime }-\mathbf{q}$ are in the vicinity of the Fermi circle.

In the next section we will derive Eliashberg equations for the model
described by Eqs.~(\ref{a01}), (\ref{a3}), (\ref{a4}), and (\ref{a8}).

\section{Eliashberg equations}

In order to describe the electron pairing mediated by electron-phonon
interaction, we derive the \emph{Eliashberg equations}~\cite%
{eliashberg,mcmillan,vonsovsky, carbotte} for the system under consideration
using conventional methods of quantum field theory.~\cite{agd} We introduce
the imaginary time normal and anomalous Green functions that are $4\times 4$
matrices in the valley and spin space
\begin{equation}
\hat{G}_{\mathbf{p}}(\tau )\equiv -\langle T_{\tau }\hat{\psi}_{\mathbf{p}%
}(0)\hat{\bar{\psi}}_{\mathbf{p}}(\tau )\rangle \quad \hat{F}_{\mathbf{p}%
}(\tau )\equiv \langle T_{\tau }\hat{\psi}_{\mathbf{-p}}(0)\hat{\psi}_{%
\mathbf{p}}^{T}(\tau )\rangle,  \label{b1}
\end{equation}%
where the field operators $\psi _{\mathbf{p}}\left( \tau \right) $ are $4$%
-components vectors having as components the operators $a_{\alpha ,\mathbf{p}%
,\sigma }$ in the Heisenberg representation.

In the isospin representation the gap function $\hat{\Delta}_{\mathbf{p}%
,\varepsilon _{n}}$ entering the Gorkov equation is a $4\times 4$ matrix.

Using the normal $\hat{G}_{\mathbf{p},\varepsilon _{n}}$ and anomalous $\hat{%
F}_{\mathbf{p},\varepsilon _{n}}$ Green functions, we consider the Dyson
equations containing both normal and anomalous self-energy parts. Using this
matrix representation, we derive the Gorkov equations in the generalized form%
\begin{eqnarray}
\left( iZ_{n}\varepsilon _{n}-\xi _{\mathbf{p}}\right) \hat{G}_{\mathbf{p,}%
\varepsilon _{n}}+Z_{n}\hat{\Delta}_{\mathbf{p},\varepsilon _{n}}F_{\mathbf{%
p,}\varepsilon _{n}}^{+} &=&1  \label{b1a} \\
\left( iZ_{n}\varepsilon _{n}+\xi _{\mathbf{p}}\right) F_{\mathbf{p,}%
\varepsilon _{n}}^{+}+Z_{n}\hat{\Delta}_{\mathbf{p},\varepsilon _{n}}^{+}%
\hat{G}_{\mathbf{p,}\varepsilon _{n}} &=&0,  \notag
\end{eqnarray}%
where $\varepsilon _{n}=\pi T\left( 2n+1\right) $ is the fermionic Matsubara
frequency, and the symbol $``+"$ stands for the Hermitian conjugation of the
$4\times 4$ matrices.

According to the Eliashberg \cite{eliashberg} theory developed for an
arbitrary value of the coupling in the limit $\omega _{D}\ll \mu $, where $%
\omega _{D}$ is the Debye frequency, one has to take into account normal
contributions of the self-energy to the Green functions, but neglect the
renormalization of the vertices. As we assume that both the electron-phonon
and electron-electron interactions are not necessarily weak, one should introduce the
factor $Z_{n}$ coming from the normal self-energy. This factor renormalizes
the coefficient in front of the frequency, but the renormalization of the
coefficient for $\xi _{\mathbf{p}}$ is neglected. This non-equivalence
originates from the fact that the renormalization of the coefficient in
front of the frequency is proportional to $\omega _{D}^{-1},$ while
corrections to the coefficient in front of $\xi _{\mathbf{p}}$ are
proportional to $\mu ^{-1}$. The factor $Z_{n}$ in front of $\hat{\Delta}_{%
\mathbf{p},\varepsilon_n}$ is written for a convenience.

By solving Eqs.~(\ref{b1a}), we write the anomalous Green function $F^{+}$ as%
\begin{equation}
\hat{F}_{\mathbf{p},\varepsilon _{n}}^{+}=\left( Z_{n}^{2}\varepsilon
_{n}^{2}+\xi _{\mathbf{p}}^{2}+Z_{n}^{2}\hat{\Delta}_{\mathbf{p}%
,\varepsilon_n}^{+}\hat{\Delta}_{\mathbf{p},\varepsilon_n}\right) ^{-1}Z_{n}%
\hat{\Delta}_{\mathbf{p},\varepsilon_n}^{+}.  \label{b1b}
\end{equation}

Explicit calculations show that, in the absence of magnetic interactions in
the system, the most favorable is singlet pairing. Therefore, we do not
present here calculations for the general case and consider only the singlet
pairing. At the same time, the structure of the gap function $\hat{\Delta}_{%
\mathbf{p},\varepsilon_n}$ can be non-trivial due to the presence of two
valleys. In our representation using the isospin, $\hat{\Delta}_{\mathbf{p}%
,\varepsilon_n}$ is a $4\times 4 $ matrix in both spin and isospin space and
we write it as
\begin{equation}
\hat{\Delta}_{\mathbf{p},\varepsilon_n}=%
\begin{pmatrix}
\Delta _{\mathbf{p},\varepsilon_n}^{11} & \Delta _{\mathbf{p}%
,\varepsilon_n}^{12} \\
\Delta _{\mathbf{p},\varepsilon_n}^{21} & \Delta _{\mathbf{p}%
,\varepsilon_n}^{22}%
\end{pmatrix}%
\otimes i\sigma _{2},  \label{b2}
\end{equation}%
where $\sigma _{2}$ is the second Pauli matrix. The fact that only this
matrix enters the gap function in the spin space is standard for the singlet
pairing. Exchanging the electrons the gap function must change the sign. As
the matrix $\sigma _{2}$ is antisymmetric, one comes to the relation
\begin{equation}
\Delta _{\mathbf{p},\varepsilon_n}^{\alpha \beta }=\Delta _{-\mathbf{p}%
,\varepsilon_n}^{\beta \alpha }  \label{b3}
\end{equation}%
Considering the triplet order parameter, one would obtain instead of the
symmetric relation for $\Delta _{\mathbf{p},\varepsilon_n}^{\alpha \beta }$
 [Eq.~(\ref{b3})] an antisymmetric one.

As the spectra of the valleys are identical, we can consider a simpler form
of Eq.~(\ref{b2}) taking%
\begin{align}
\Delta _{\mathbf{p},\varepsilon_n}^{11}&=\Delta _{\mathbf{p}%
,\varepsilon_n}^{22}=\Delta _{0,\mathbf{p},\varepsilon_n}=\Delta _{0,-%
\mathbf{p},\varepsilon_n},\text{ }  \notag \\
\Delta _{\mathbf{p},\varepsilon_n}^{12}&=\Delta _{-\mathbf{p}%
,\varepsilon_n}^{21}=\Delta _{\mathbf{Q,p},\varepsilon_n} .  \label{b4}
\end{align}

The gap $\Delta _{0,\mathbf{p},\varepsilon_n}$ corresponds to the
intravalley pairing and $\Delta _{\mathbf{Q,p},\varepsilon_n}$ to the
intervalley one.

By substituting Eqs.~(\ref{b2}) and (\ref{b4}) into (\ref{b1b}), we write the
solution for the anomalous Green function $\hat{F}_{\mathbf{p}%
,\varepsilon_n} $ as

\begin{align}
\hat{F}_{\mathbf{p},\varepsilon _{n}}=\frac{Z_{n}(P_{\mathbf{p},\varepsilon
_{n}}^{+}+P_{\mathbf{p},\varepsilon _{n}}^{-})}{2}%
\begin{pmatrix}
\Delta _{0,\mathbf{p},\varepsilon_n} & \Delta _{\mathbf{Q,p},\varepsilon_n}
\\
\Delta _{\mathbf{Q,-p},\varepsilon_n} & \Delta _{0,\mathbf{p},\varepsilon_n}%
\end{pmatrix}%
& \otimes i\sigma _{2}  \notag \\
+\frac{Z_{n}(P_{\mathbf{p},\varepsilon _{n}}^{+}-P_{\mathbf{p},\varepsilon
_{n}}^{-})}{2}%
\begin{pmatrix}
\Delta _{\mathbf{Q,-p},\varepsilon_n} & \Delta _{0,\mathbf{p},\varepsilon_n}
\\
\Delta _{0,\mathbf{p},\varepsilon_n} & \Delta _{\mathbf{Q},\mathbf{p}%
,\varepsilon_n}%
\end{pmatrix}%
& \otimes i\sigma _{2},  \label{F}
\end{align}%
where the function $P_{\mathbf{p},\varepsilon _{n}}^{\pm }$ equals
\begin{equation}
P_{\mathbf{p},\varepsilon _{n}}^{\pm }=\frac{1}{(Z_{n}\varepsilon_n)^{2}+\xi
_{\mathbf{p}}^{2}+Z_{n}^{2}\left\vert \Delta _{0,\mathbf{p}%
,\varepsilon_n}\pm \Delta _{\mathbf{Q,p},\varepsilon_n}\right\vert ^{2}} .
\label{b5}
\end{equation}

Equation~(\ref{F}) should be complemented by a self-consistency equation, which
is actually a matrix equation in the isospin space. Writing separately the
anomalous and normal parts of the self-energy, we come to the following
equations:
\begin{widetext}
\begin{align}
Z_n\hat \Delta_{\mathbf{p},\varepsilon_n} =
 &
 \left( T\sum_{j,m}\int \frac{ d \mathbf{p}^\prime}{(2\pi)^2} D_j(\mathbf{q},\varepsilon_n-\varepsilon_m) \left|g^{11}_{\bp,\bq,j} \right|^2
+ T\sum_{m}\int \frac{d\mathbf{p}^\prime}{(2\pi)^2} V_{\bp\bpp}
\right)\begin{pmatrix}
                       F_{\mathbf{p'},\varepsilon_{m}}^{11}  & F_{\mathbf{p'},\varepsilon_{m}}^{12}\\
                       F_{\mathbf{p'},\varepsilon_{m}}^{21} &  F_{\mathbf{p'},\varepsilon_{m}}^{22}
                       \end{pmatrix}
                       \nonumber\\
        +
        &
        \left( T\sum_{j,m,\pm}\int \frac{d \mathbf{p}^\prime}{(2\pi)^2} \frac 12 D_j(\mathbf{q \pm Q},\varepsilon_n-\varepsilon_m)         \left|g^{12}_{\bp,\bq,j}\right|^2
+ T\sum_{m,\pm}\int \frac{d\mathbf{p}^\prime}{(2\pi)^2} \frac 12 V_{\mathbf{p p^\prime \pm Q}}
\right)
\begin{pmatrix}
                       0  & F_{\mathbf{p'},\varepsilon_{m}}^{12}\\
                       F_{\mathbf{p'},\varepsilon_{m}}^{21} &  0
                       \end{pmatrix},
                       \label{dyson}
\\
(1-Z_n) \ii \varepsilon_n \hat{\mathds{1}}
=
&
\left( T\sum_{j,m}\int \frac{d\mathbf{p}^\prime}{(2\pi)^2} D_j(\mathbf{q},\varepsilon_n-\varepsilon_m) \left|g^{11}_{\bp,\bq,j} \right|^2
+ T\sum_{m}\int \frac{d \mathbf{p}^\prime}{(2\pi)^2} V_{\mathbf{p p^\prime}}\right.
\nonumber\\
&\left. +T\sum_{j,m,\pm}\int \frac{d\mathbf{p}^\prime}{(2\pi)^2} \frac 12 D_j(\mathbf{q \pm Q},\varepsilon_n-\varepsilon_m)         \left|g^{12}_{\bp,\bq,j}\right|^2
+ T\sum_{m,\pm}\int \frac{d\mathbf{p}^\prime}{(2\pi)^2} \frac 12 V_{\mathbf{p p^\prime \pm Q}}
\right)
\begin{pmatrix}
                       G_{\mathbf{p'},\varepsilon_{m}}^{11}  & 0\\
                       0 &  G_{\mathbf{p'},\varepsilon_{m}}^{22}
                       \end{pmatrix},
                       \label{b6}
\end{align}
\end{widetext}

where $D_{j}$ is the phonon Green function for the polarization $j$,
\begin{equation}
D_j\left( \mathbf{q,}\omega _{n}\right) =-\frac{2\omega _{j}\left( \mathbf{q}%
\right) }{\omega _{n}^{2}+\omega _{j}^{2}\left( \mathbf{q}\right) }.
\label{b6a}
\end{equation}

In Eqs.~(\ref{dyson}) and  (\ref{b6}), $\mathbf{q}=\mathbf{p}-\mathbf{p}^{\prime }$
and the symmetry relation $\left\vert g^{\alpha \beta }\right\vert
^{2}=\left\vert g^{\beta \alpha }\right\vert ^{2}$ for the coupling
functions is used. Further, we neglect off-diagonal terms of the normal
Green function.

Equation~(\ref{b6}) describes the normal contribution to the self-energy.
The intravalley and intervalley scattering contributions enter on equal footing.
In principle, the integrals in the right-hand side contain not only linear in $%
\varepsilon _{n}$ contributions, but also renormalize the Fermi energy and
the spectrum. The latter types of the contributions do not lead to important
changes of physical quantities and are neglected.

Equation~(\ref{dyson}) is the self-consistency equation for the order
parameter $\hat{\Delta}_{\mathbf{p},\varepsilon_n}$. It is clear that the
equation for the intravalley order parameter $\Delta^{intra}_{\mathbf{p}%
,\varepsilon_n}=\Delta _{0,\mathbf{p},\varepsilon_n}$ (diagonal elements of
the matrices in Eq.~(\ref{dyson})) differs from the one for the intervalley
gap function $\Delta^{inter}_{\mathbf{p},\varepsilon_n}=\Delta _{\mathbf{Q,p}%
,\varepsilon_n}$ (off diagonal elements of the matrices in Eq.~(\ref{dyson}%
)). For the former, only the intravalley interaction is important, while for
the latter, both the intravalley and intervalley interactions contribute. It is
clear that, provided the intervalley interaction is negative, the
intervalley pairing is more favorable than the intravalley one.

For explicit calculations, it is convenient to use the representation of the
temperature Green functions in terms of retarded Green's functions
\begin{align}
\hat{F}_{\mathbf{p},\varepsilon _{n}}& =\int_{-\infty }^{\infty }\frac{dz}{%
2\pi }\frac{2\mathrm{Im}\hat{F}^{R}(\mathbf{p},z)}{z-i\varepsilon _{n}},
\label{b7} \\
D_{j,{\mathbf{q},\Omega_{n}}}& =\int_{-\infty }^{\infty }\frac{dz}{2\pi }%
\frac{b_{j}(\mathbf{q},z)}{z-i\Omega _{n}},  \label{b8}
\end{align}%
where $\hat{F}^{R}$ is the retarded Green function, and $b_{j}=2\mathrm{Im}%
D_{j}^{R}$ is the phonon spectral function.

Substituting Eqs.~(\ref{b7}, \ref{b8}) into Eq.~(\ref{dyson}), we carry out
the summation over the Matsubara frequencies and perform an analytic
continuation $i\varepsilon _{n}\rightarrow \omega +i\delta $ with an
infinitesimal positive $\delta $.

Considering first the contribution of the intravalley pairing to the gap
function (first line of Eq.~(\ref{dyson})) we bring it to the form
\begin{align}
\hat{\Delta}_{\mathbf{p,}\omega }^{intra}& =Z^{-1}\left( \omega \right) \int
\frac{d\mathbf{p}^{\prime }}{(2\pi )^{2}}\int_{-\infty }^{\infty }\frac{dz}{%
2\pi }\int_{-\infty }^{\infty }\frac{dz^{\prime }}{2\pi }\ b_{j}(\mathbf{p-p}%
^{\prime },z)  \notag \\
& \times \left\vert g_{\mathbf{p,q},j}^{11}\right\vert ^{2}\frac{\tanh \frac{%
z^{\prime }}{2T}+\coth \frac{z}{2T}}{\omega -z-z^{\prime }+i\delta }\mathrm{%
Im}\hat{F}^{R}(\mathbf{p^{\prime }},z)  \notag \\
& +\int \frac{d\mathbf{p}^{\prime }}{(2\pi )^{2}}\int_{-\infty }^{\infty }%
\frac{dz}{2\pi }\ V_{\mathbf{p}\mathbf{p}^{\prime }}\tanh \frac{z}{2T}%
\mathrm{Im}\hat{F}^{R}(\mathbf{p^{\prime }},z).  \label{dysonsum}
\end{align}%
The contribution $\hat{\Delta}_{\mathbf{p},\omega }^{inter}$ to the gap
function coming from the intervalley pairing can be written similarly.

Now, we will analyze the phonon and Coulomb parts separately.

\subsection{Phonon part}

In the case of large doping levels considered here one can use standard
approximations well known in the theory of conventional metals. In particular,
only momenta close to the Fermi surface can be taken into account. Reducing
the dependence on the momenta $\mathbf{p}$ by the dependence on the unit
vector $\mathbf{n,}$ $\mathbf{p=}p_{F}\mathbf{n}$, we average the gap
function $\hat{\Delta}_{\mathbf{p,}\varepsilon }$ over the Fermi~surface and
introduce the quantity
\begin{equation}
\hat{\Delta}_{\omega }=\int_{S_{F}}d\mathbf{n}\ \hat{\Delta}_{p_{F}\mathbf{n,%
}\omega },  \label{b9}
\end{equation}%
where $\int_{S_{F}}d\mathbf{n}$ denotes the integral over all directions on
the Fermi surface, and $\nu $ is the density of states at the Fermi energy.
The normalization is chosen in such a way that
\begin{equation}
\int_{S_{F}}d\mathbf{n}=1.  \label{b10}
\end{equation}%
The integral over the momentum in the right-hand side of Eq.~(\ref{dysonsum}) reduces
in this approximation to the form
\begin{equation}
\int \left( ...\right) \frac{d^{2}\mathbf{p}}{(2\pi )^{2}}=\nu \left( \mu
\right) \int_{S_{F}}d\mathbf{n}\int_{-\infty }^{\infty }d\xi _{\mathbf{p}%
}\left( ...\right)  \label{b11}
\end{equation}%
where $\nu \left( \mu \right) $ is the density of states, Eq.~(\ref{DOS}).

Using for the phonon Green function its bare value, Eq.~(\ref{b6a}), such
that
\begin{equation}
b_{j}\left( \mathbf{q},z\right) =2\pi \left[ \delta (z+\omega _{j}(\mathbf{q}%
))-\delta (z-\omega _{j}(\mathbf{q}))\right]  \label{b12}
\end{equation}%
and integrating over the variable $\xi _{\mathbf{p}}$, we write the phonon
contribution $\left( \hat{\Delta}_{\omega }^{intra}\right) _{ph}$ to the gap
function $\hat{\Delta}_{\omega }^{intra}$ of Eq.~(\ref{dysonsum}) as
\begin{equation}
\left( \Delta _{\omega }^{intra}\right) _{ph}=Z^{-1}\left( \omega \right)
\int_{-\infty }^{\infty }\ K_{ph}^{11}(z,\omega )\ \mathrm{Im}\left( \bar{F}%
^{R}(z)\right) ^{11}dz.  \label{b12a}
\end{equation}%
The intravalley phonon kernel $K_{ph}^{11}(z,\omega )$ entering Eq.~(\ref%
{b12a}) can be written as
\begin{align}
& K_{ph}^{11}(z,\omega )=-\frac{1}{2}\int_{0}^{\infty }dz^{\prime }\ \alpha
_{11}^{2}f(z^{\prime })  \notag \\
\times & \left( \frac{\tanh \frac{z}{2T}+\coth \frac{z^{\prime }}{2T}}{%
\omega -z^{\prime }-z+i\delta }-\frac{\tanh \frac{z}{2T}-\coth \frac{%
z^{\prime }}{2T}}{\omega +z^{\prime }-z+i\delta }\right) ,  \label{kernel}
\end{align}%
where $\alpha _{11}^{2}f(z)$ is the \emph{Eliashberg function} for
intravalley phonon scattering processes
\begin{align}
\alpha _{11}^{2}f(z)=\nu \int_{S_{F}}d\mathbf{n}& \int_{S_{F}}d\mathbf{n}%
^{\prime }\sum_{j}\left\vert g_{p_{F}\mathbf{n,}p_{F}\left( \mathbf{n-n}%
^{\prime }\right) }^{11}\right\vert ^{2}  \notag \\
& \times \delta (z-\omega _{j}(p_{F}\left( \mathbf{n-n}^{\prime }\right) )).
\label{b14}
\end{align}

According to Eq.~(\ref{b14}) this function contains the double average over
the Fermi surface of the electron-phonon coupling function squared.

The function $\bar{F}(z)$ in Eq.~(\ref{b12a}) equals
\begin{equation}
\bar{F}^{R}=\int_{-\infty }^{\infty }\hat{F}^{R}d\xi ,  \label{b13}
\end{equation}%
where $\hat{F}^{R}$ is the retarded anomalous Green function obtained from
the corresponding temperature Green function, Eq.~(\ref{F}).

This integration over $\xi $ in Eq.~(\ref{b13}) results in a replacement of
the functions $P_{\pm }$ by $\bar{P}_{\pm }$ given by
\begin{equation}
\bar{P}_{\pm }(z)=\frac{i\pi \ \mathrm{sign}(z)}{Z(z)\sqrt{z^{2}-\left\vert
\Delta _{0,\mathbf{p},z}\pm \Delta _{\mathbf{Q,p},z}\right\vert ^{2}}}.
\label{b15}
\end{equation}%
As concerns the off-diagonal parts of the gap function, we have to include
the intervalley interaction processes into the self-consistency relation.
Then we obtain
\begin{align}
\left( \Delta_{\omega }^{inter} \right) _{ph}=Z^{-1}\left( \omega \right)
\int_{-\infty }^{\infty }& \ K_{ph}^{11}(z,\omega )\ \mathrm{Im}\left( \bar{F%
}^{R}(z)\right) ^{12}dz  \notag \\
+Z^{-1}\left( \omega \right) \int_{-\infty }^{\infty }& \
K_{ph}^{12}(z,\omega )\ \mathrm{Im}\left( \bar{F}^{R}(z)\right) ^{12}dz,
\label{gapphQ}
\end{align}%
with the intervalley kernel
\begin{align}
& K_{ph}^{12}(z,\omega )=-\frac{1}{2}\int_{0}^{\infty }dz^{\prime }\ \alpha
_{12}^{2}f(z^{\prime })  \notag \\
\times & \left( \frac{\tanh \frac{z}{2T}+\coth \frac{z^{\prime }}{2T}}{%
\omega -z^{\prime }-z+i\delta }-\frac{\tanh \frac{z}{2T}-\coth \frac{%
z^{\prime }}{2T}}{\omega +z^{\prime }-z+i\delta }\right) ,  \label{kerneli}
\end{align}%
where the Eliashberg function for the intervalley scattering equals
\begin{align}
\alpha _{12}^{2}f(z)=\nu \int_{S_{F}}d\mathbf{n}& \int_{S_{F}}d\mathbf{n}%
^{\prime }\sum_{j,\pm } \frac 12 \left\vert g_{p_{F}\mathbf{n,}p_{F}\left(
\mathbf{n-n}^{\prime }\right) }^{12}\right\vert ^{2}  \notag \\
\times & \delta (z-\omega _{j}(p_{F}\left( \mathbf{n-n}^{\prime }\right) \pm
\mathbf{Q})).  \label{b16}
\end{align}%
Similar calculations for the normal self-energy part lead to an expression
applicable for the contribution of both inter- and intravalley phonon modes
\begin{align}
& (1-Z(\omega ))\omega  \label{normal} \\
& =\int_{-\infty }^{\infty }(K_{ph}^{11}(z,\omega )+K_{ph}^{12}(z,\omega ))%
\mathrm{Im}\left( \bar{G}^{R}(z)\right) dz,  \notag
\end{align}%
where
\begin{equation}
\bar{G}^{R}=\int_{-\infty }^{\infty }\hat{G}^{R}d\xi ,  \label{b18}
\end{equation}%
The formulas derived in this section completely describe the effects of the
electron-phonon interactions. Now, we will investigate the remaining parts
of Eq.~(\ref{dyson}) originating from the Coulomb interaction.

\subsection{Coulomb part}

Calculating the Coulomb part in Eq.~(\ref{dyson}) one should first
renormalize the Coulomb interaction integrating out high energy degrees of
freedom \cite{morel,vonsovsky} and thus reduce the original model to a model
with a certain energy cutoff $\omega _{c},$ such that $\omega _{D}\ll \omega
_{c}\ll \mu $, where $\omega _{D}=2300 \ \mathrm{K}$. This renormalization
is also logarithmic and can easily be carried out in the ladder
approximation. The final results can be expressed in terms of the $\emph{%
pseudopotentials}$ $U^{11}$ and $U^{12}$, respectively
\begin{align}
U^{11}& =\frac{V^{11}}{1+\nu V^{11}\ln (\frac{\mu }{\omega _{c}})},  \notag
\\
U^{12}& =\frac{V^{11}+V^{12}}{1+\nu (V^{11}+V^{12})\ln (\frac{\mu }{\omega
_{c}})}  \label{Coulomb}
\end{align}%
where%
\begin{equation}
V^{\alpha \beta }=\int_{S_F} \int_{S_{F}}V_{p_{F} \left( \mathbf{n-n}%
^{\prime }\right)}^{\alpha \beta }d\mathbf{n} \ d\mathbf{n}^{\prime }
\label{b19}
\end{equation}%
and the matrix elements $V_{\mathbf{q}}^{11}$, $V_{\mathbf{q}}^{12}$ are
given by (\ref{a9}, \ref{a7}).

Equations~(\ref{Coulomb}) show that the effective Coulomb interaction $U_{\mathbf{%
q}}^{\alpha \beta }$ can not be very strong. Since the Coulomb potential
decays as $\left\vert \mathbf{q}\right\vert ^{-1}$ in the momentum space,
the function $V_{\mathbf{q}}^{12}$ is smaller than $V_{\mathbf{q}}^{11}$ and
the pseudopotentials do not differ much from each other. The
pseudopotentials $U_{\mathbf{q}}^{\alpha \beta }$ monotonically grow with
increasing $V^{\alpha \beta }$, but their values are limited by $\nu ^{-1} $.

The renormalization of the Coulomb interaction (\ref{Coulomb}) is not
important near the Dirac point because both $\nu $ and $\mu $ are small, but
can considerably reduce it in the region $\mu \gtrsim 1\mathrm{eV}$.

As concerns the normal self-energy, a contribution coming from the Coulomb
interaction is small and can be neglected.~\cite{vonsovsky}

Using the pseudopotentials $U^{11}$ and $U^{12}$ [Eqs.~(\ref{Coulomb})], we
write their contributions $\Delta _{C}^{11}$ and $\Delta _{C}^{12}$ to the
gap function as

\begin{align}
\left( \Delta_{\omega }^{intra}\right) _{C}& =\ \nu Z^{-1}\left( \omega
\right) U^{11}\ \int\limits_{0}^{\omega _{c}}\tanh \frac{z}{2T}\ \mathrm{Im}%
\left( \bar{F}^{R}(z)\right) ^{11}\frac{dz}{2\pi },  \label{gapc} \\
\left( \Delta_{\omega }^{inter}\right) _{C}& =\ \nu Z^{-1}\left( \omega
\right) U^{12}\int\limits_{0}^{\omega _{c}}\tanh \frac{z}{2T}\ \mathrm{Im}%
\left( \bar{F}^{R}(z)\right) ^{12}\frac{dz}{2\pi }.  \label{gapcQ}
\end{align}

Equations (\ref{b12a}), (\ref{gapphQ}), (\ref{gapc}), and (\ref{gapcQ}) are
basic equations of the electron-phonon theory of superconductivity in
graphene. Based on them, we can derive a formula for the critical
temperature.

\subsection{Critical temperature}

In order to make an estimate for the critical temperature, we simplify our
equations according to standard procedures. Our goal is to clarify which
type of  pairing is more favorable and to estimate the critical
temperature rather than to calculate it from first principles. Within
these procedures, the calculations become considerably simpler, but we believe
that our goal is still achieved.

Following Ref.~[\onlinecite{vonsovsky}], we approximate the system of
equations by linearizing their right-hand side with respect to the gap functions, and
we approximate the phonon kernels Eqs.~(\ref{kernel}) and  (\ref{kerneli}) by the
following expressions:
\begin{equation}
K_{ph}^{\alpha \beta }(z,\omega )=%
\begin{cases}
\frac{\lambda _{\alpha \beta }}{2}\tanh \frac{z}{2T_{c}} & \quad \left\vert
z\right\vert ,\left\vert \omega \right\vert <\omega _{D} \\
0 & \quad \mathrm{otherwise}%
\end{cases}
\label{b21}
\end{equation}%
where $\lambda _{\alpha \beta }$, $(\alpha ,\beta =1,2)$ are the intravalley and
intervalley electron-phonon coupling constants
\begin{equation}
\lambda _{\alpha \beta }=2\int_{0}^{\infty }\frac{\alpha _{\alpha \beta
}^{2}f(z)}{z}dz.  \label{b22}
\end{equation}%
(Actually, we have to calculate only $\lambda _{11}$ and $\lambda _{12}$
because $\lambda _{22}=\lambda _{11}$ and $\lambda _{12}=\lambda _{21}.$)
Equations~(\ref{b21}) show that the further calculations can be performed
independently for the quantities with $``11"$and $``12"$.

The function $Z(\omega)$ coming from the normal self-energy, Eq.~(\ref%
{normal}), is just a constant and can be written as
\begin{equation}
Z=1+\lambda _{11}+\lambda _{12}\equiv 1+\lambda .  \label{b23}
\end{equation}

Then, the fact that $Z$ does not depend on frequency leads us to the
conclusion that the gap function $\Delta _{\alpha \beta }$ does not depend
on the frequency for $\omega <\omega _{D}$ either [see Eqs.~(\ref{gapc}) and (\ref%
{gapcQ})].

By using Eqs.~(\ref{b21}) and linearizing the self-consistency equations (\ref%
{b12a}), (\ref{gapphQ}), (\ref{gapc}), and (\ref{gapcQ}), we can derive equations for
the critical temperatures of the intravalley and intervalley pairings $%
T_{c}^{intra}$ and $T_{c}^{inter}$, respectively. At the end, only the
pairing with a higher critical temperature should be kept and used for the
description of the superconductivity.

In order to obtain the equation for the critical temperature we choose the
following form of the function $\Delta _{\omega }$:
\begin{equation}
\Delta _{\omega }=%
\begin{cases}
\Delta _{ph}\qquad \omega <\omega _{D} \\
\Delta _{C}\qquad \omega _{D}<\omega <\omega _{c}%
\end{cases}
\label{b23a}
\end{equation}%
with constants $\Delta _{ph}$ and $\Delta _{C}$.

Using the approximation, Eq.~(\ref{b23a}), we reduce the equation for the
critical temperature $T_{c}^{intra}$ to the form
\begin{equation}
\int_{0}^{\omega _{D}}\frac{dz}{z}\tanh \frac{z}{2T_{c}^{intra}}=\frac{%
1+\lambda _{11}+\lambda _{12}}{\lambda _{11}-\mu _{11}^{\ast }},  \label{b24}
\end{equation}%
where the parameter $\mu _{11}^{\ast }$ equals
\begin{equation}
\mu _{11}^{\ast }=\frac{\nu V_{11}}{1+\nu V_{11}\ln (\frac{\mu }{\omega _{D}}%
)}.  \label{b25}
\end{equation}%
Note that the approximation written in Eq.~(\ref{b23a}) leads to the
replacement of $\omega _{c}$ in the argument of the logarithm in Eq.~(\ref%
{Coulomb}) by the Debye frequency $\omega _{D}$ in Eq.~(\ref{b25}).

The solution of Eq.~(\ref{b24}) exists only when the right-hand side is positive.
Therefore, the intravalley pairing is possible provided $\lambda _{11}>\mu
_{11}^{\ast }$.

By calculating the integral over $z$ in Eq.~(\ref{b24}), we write the critical
temperature $T_{c}^{intra}$ of the intravalley pairing explicitly%
\begin{equation}
T_{c}^{intra}=1{.}13\omega _{D}\ \exp \left( -\frac{1+\lambda }{\lambda
_{11}-\mu _{11}^{\ast }}\right) .  \label{b26}
\end{equation}

As concerns the intervalley pairing, both the intra- and intervalley phonon
interactions contribute and we come to the following equation for the
critical temperature $T_{c}^{inter}$ of the intervalley pairing
\begin{equation}
\int_{0}^{\omega _{D}}\frac{dz}{z}\tanh \frac{z}{2T^{inter}_{c}}=\frac{%
1+\lambda _{11}+\lambda _{12}}{\lambda _{11}+\lambda _{12}-\mu _{12}^{\ast }}%
,  \label{b27}
\end{equation}%
with the renormalized Coulomb interaction given by
\begin{equation}
\mu _{12}^{\ast }=\frac{\nu (V_{11}+V_{12})}{1+\nu (V_{11}+V_{12})\ln (\frac{%
\mu }{\omega _{D}})}.  \label{pseudopotential}
\end{equation}

The intervalley superconductivity is possible provided the condition $%
\lambda =\lambda _{11}+\lambda _{12}>\mu _{12}^{\ast }$ is fulfilled and we
obtain the critical temperature $T_{c}^{inter}$ for this type of the
superconductivity in the form
\begin{equation}
T_{c}^{inter}=1{.}13\omega _{D}\ \exp \left( -\frac{1+\lambda }{\lambda -\mu
_{12}^{\ast }}\right) .  \label{Tc}
\end{equation}

As the constant $\mu _{12}^{\ast }$ only slightly exceeds $\mu _{11}^{\ast }$
for a strongly renormalized Coulomb interaction, the intervalley pairing
looks more favorable. Moreover, the electron-phonon coupling $\lambda $
entering Eq.~(\ref{Tc}) is a quantity that can be extracted directly from
the angle-resolved photoemission spectroscopy (ARPES), which simplifies
estimates of the transition temperature $T_{c}^{inter}$, Eq.~(\ref{Tc}).
Clearly, the superconductivity is possible provided the condition
\begin{equation}
\lambda >\mu _{12}^{\ast }  \label{b28}
\end{equation}%
is fulfilled. Explicit estimates for the critical temperature $T_{c}^{inter}$
are performed in Sec.~$\mathrm{V}$. We restrict ourselves with calculation
of the temperature $T_{c}^{inter}$ because the critical temperature of the
intravalley coupling $T_{c}^{intra}$, Eq.~(\ref{b26}) is lower than $%
T_{c}^{inter}$ for realistic parameters of $\mu _{11}^{\ast }$ and $\mu
_{12}^{\ast }$ . In addition, the intravalley pairing is sensitive to
impurity scattering, which contrasts the intervalley pairing. The effect of
the impurities on the two types of the superconducting pairings is
considered in the next section.

\section{Impurities}

In the previous sections, we considered superconductivity in clean systems.
Usually, it is assumed that non-magnetic impurities do not affect the
superconducting transition temperature.~\cite{agd} However, the situation is
not as simple for a system with several valleys, where some of the
superconducting correlations can be sensitive to the impurities. We have
considered the superconducting intervalley and intravalley pairing in the clean
graphene, and now we will study effects of the potential impurities on these
types of the superconductivity.

In order to model this, we introduce an impurity Hamiltonian for the
two-valley system. Generally, both intervalley and intravalley impurity
scatterings are possible. The most general form of the Hamiltonian for
disordered graphene taking into account its Dirac-type spectrum has been
written (in absence of electron-electron interactions) in Ref.~[%
\onlinecite{aleiner}]. However, as we consider graphene for energies far
away for the Dirac point, we introduce a standard impurity Hamiltonian in
the momentum space
\begin{equation}
\hat{H}_{imp}=\sum_{\mathbf{p},\mathbf{q},\sigma }u_{\mathbf{q}}c_{\mathbf{p}%
+\mathbf{q},\sigma }^{\dag }c_{\mathbf{q,}\sigma }  \label{c1}
\end{equation}%
with the momentum-dependent impurity potential $u_{\mathbf{q}}$. By using the
isospin representation, we rewrite this expression in the form
\begin{equation}
\hat{H}_{imp}=\sum_{\alpha ,\beta =1}^{2}\sum_{\mathbf{p},\mathbf{q},\sigma
}u_{\mathbf{q}}^{\alpha \beta }a_{\alpha ,\mathbf{p}+\mathbf{q},\sigma
}^{\dag }a_{\beta ,\mathbf{p},\sigma },  \label{c2}
\end{equation}%
where the functions $u_{\mathbf{q}}^{\alpha \beta }$ are related to the
scattering potential as
\begin{equation*}
u_{\mathbf{q}}^{11}=u_{\mathbf{q}}^{22}=u_{\mathbf{q}},\quad u_{\mathbf{q}%
}^{12}=u_{\mathbf{q}-\mathbf{Q}},\quad u_{\mathbf{q}}^{21}=u_{\mathbf{q}+%
\mathbf{Q}}.
\end{equation*}

Since intervalley scattering processes require a large momentum transfer,
they can not be caused by Coulomb impurities of the substrate. On the other
hand, vacancies in the graphene sheet, adatoms, surface ripples, or
topological defects can lead to both intravalley and intervalley scattering
events.~\cite{castro}

For calculations, we use the standard diagrammatic approach and treat the
corrections in the Born approximation.~\cite{agd, bennemanng} Studying the
system far from the Dirac points, we consider only diagrams with
non-crossing impurity lines. For the calculations, we assume that the
effects of electron-phonon and Coulomb interaction have already been taken
into account according to Eqs.~(\ref{b1a}), (\ref{dyson}), and (\ref{b6}), which
determines $Z_{n}$ and $\hat{\Delta}$. Calculating the corrections to these
quantities arising from the impurity scattering we denote the renormalized
values by $\tilde{Z}_{n}$ and $\tilde{\Delta}$, respectively.

By using the standard diagrammatic expansion in the approximation of
non-crossing impurity lines, we obtain the modified self-energy equations
\begin{align}
\tilde{Z}_{n}\tilde{\hat{\Delta}}_{\mathbf{p},\varepsilon _{n}}-Z_{n}\hat{%
\Delta}_{\mathbf{p},\varepsilon _{n}}& =\int \frac{d\mathbf{p}^{\prime }}{%
(2\pi )^{2}}\left\vert u_{\mathbf{q}}^{11}\right\vert ^{2}%
\begin{pmatrix}
F_{\mathbf{p^{\prime }},\varepsilon _{m}}^{11} & F_{\mathbf{p^{\prime }}%
,\varepsilon _{m}}^{12} \\
F_{\mathbf{p^{\prime }},\varepsilon _{m}}^{21} & F_{\mathbf{p^{\prime }}%
,\varepsilon _{m}}^{22}%
\end{pmatrix}
\notag \\
& +\int \frac{d\mathbf{p}^{\prime }}{(2\pi )^{2}}\left\vert u_{\mathbf{q}%
}^{12}\right\vert ^{2}%
\begin{pmatrix}
0 & F_{\mathbf{p^{\prime }},\varepsilon _{m}}^{12} \\
F_{\mathbf{p^{\prime }},\varepsilon _{m}}^{21} & 0%
\end{pmatrix}
\label{imp1} \\
(Z_{n}-\tilde{Z}_{n})i\varepsilon _{n}& =\int \frac{d\mathbf{p}^{\prime }}{%
(2\pi )^{2}}\left( \left\vert u_{\mathbf{q}}^{11}\right\vert ^{2}+\left\vert
u_{\mathbf{q}}^{12}\right\vert ^{2}\right)  \notag \\
& \times
\begin{pmatrix}
G_{\mathbf{p^{\prime }},\varepsilon _{m}}^{11} & 0 \\
0 & G_{\mathbf{p^{\prime }},\varepsilon _{m}}^{22}%
\end{pmatrix}%
,  \label{imp2}
\end{align}%
where $F_{\mathbf{p},\varepsilon _{n}}$ and $G_{\mathbf{p},\varepsilon _{n}}$
denote the renormalized Green functions. To obtain these functions, one has
just to replace $Z_{n}$ and $\Delta $ by $\tilde{Z}_{n}$ and $\tilde{\Delta}$
in Eq.~(\ref{b1a}). The further calculations are similar to those performed
previously. We calculate the momentum integral in Eqs.~(\ref{imp1}) and (\ref%
{imp2}) in the standard way and expand the right-hand sides of the equations
in the gap functions $\tilde{\Delta}_{0,\mathbf{p}}$, $\tilde{\Delta}_{%
\mathbf{Q},\mathbf{p}}$, which gives us the possibility to calculate the
critical temperature $T_{c}$. As before, the intervalley interactions affect
only the intervalley gap and the normal self-energy.

This leads to the following set of equations:
\begin{align}
\tilde{Z}_{n}=& Z_{n}+\frac{1}{2\tau }\frac{1}{\sqrt{\varepsilon
_{n}^{2}+\left\vert \tilde{\Delta}_{\mathbf{Q},\mathbf{p}}+\tilde{\Delta}_{0,%
\mathbf{p}}\right\vert ^{2}}},  \label{imp1b} \\
\tilde{Z_{n}}\tilde{\Delta}_{\mathbf{Q},\mathbf{p}}=& Z_{n}\Delta _{\mathbf{Q%
},\mathbf{p}}+\frac{1}{2\tau }\frac{\tilde{\Delta}_{\mathbf{Q},\mathbf{p}}}{%
\sqrt{\varepsilon _{n}^{2}+\left\vert \tilde{\Delta}_{\mathbf{Q},\mathbf{p}}+%
\tilde{\Delta}_{0,\mathbf{p}}\right\vert ^{2}}},  \label{imp2b} \\
\tilde{Z_{n}}\tilde{\Delta}_{0,\mathbf{p}}=& Z_{n}\Delta _{0,\mathbf{p}}+%
\frac{1}{2\tau _{11}}\frac{\tilde{\Delta}_{0,\mathbf{p}}}{\sqrt{\varepsilon
_{n}^{2}+\left\vert \tilde{\Delta}_{\mathbf{Q},\mathbf{p}}+\tilde{\Delta}_{0,%
\mathbf{p}}\right\vert ^{2}}}.  \label{imp2c}
\end{align}%
Here, we have defined the elastic scattering time
\begin{equation}
\tau ^{-1}\equiv \tau _{11}^{-1}+\tau _{12}^{-1}  \label{c5}
\end{equation}%
In Eq.~(\ref{c5}), $\tau _{11}$ and $\tau _{12}$ are intravalley and intervalley
scattering times
\begin{align}
\tau _{11}^{-1}& =n_{imp}\ \nu \int_{S_{F}}d\mathbf{n}\left\vert u_{p_{F}%
\mathbf{n}}^{11}\right\vert ^{2},  \label{c6} \\
\tau _{12}^{-1}& =n_{imp}\ \nu \int_{S_{F}}d\mathbf{n}\left\vert u_{p_{F}%
\mathbf{n}}^{12}\right\vert ^{2},  \label{c7}
\end{align}%
where $n_{imp}$ is the impurity concentration. Deriving Eqs.~(\ref{imp1b})-(%
\ref{c7}), we assumed as usual that the disorder is weak. Therefore, the main
contribution in the integral over the momenta comes from the vicinity of the
Fermi energy.

Calculating $T_{c}^{inter}$ we can put in Eqs.~(\ref{imp1b}-\ref{imp2c}) $%
\tilde{\Delta}_{0,\mathbf{p}}=\Delta _{0\mathbf{,p}}=0$, which immediately
leads to the relation
\begin{equation}
\Delta _{\mathbf{Q,p}}=\tilde{\Delta}_{\mathbf{Q,p}}  \label{c8}
\end{equation}
because the normal and anomalous self energy renormalizations of $\tilde{%
\Delta}_{\mathbf{Q,p}}$ cancel each other. Using Eq.~(\ref{c8}) we conclude
that both $\Delta _{\mathbf{Q,p}}$ and $\tilde{\Delta}_{\mathbf{Q,p}}$ must
turn to zero at the same temperature $T_{c}^{inter}$ and this means the
superconducting transition temperature $T_{c}^{inter}$ for the intervalley
pairing is not affected by the disorder.

One can also come to this result by replacing the functions $Z_{n}$ and $%
\Delta _{\mathbf{Q,p}}$ in Eqs.~(\ref{b1b}) and  (\ref{F}) by $\tilde{Z}_{n}$ and $%
\tilde{\Delta}_{\mathbf{Q,p}}$ and using again Eqs.~(\ref{imp1b})-(\ref{imp2c}%
). Then, one comes to Eqs.~(\ref{gapcQ}) with $\Delta _{\mathbf{Q,p}}$
replaced by $\tilde{\Delta}_{\mathbf{Q,p}},$ which leads to Eq.~(\ref{Tc}).

On the other hand, we see that the cancelation of the normal and anomalous
self-energies does not occur when calculating $\Delta _{0,\mathbf{p}}$ at
zero $\Delta _{\mathbf{Q,p}}$, which indicates that impurities influence
this parameter. In fact, comparing this result with the conventional theory
of paramagnetic impurities in superconductors,\cite{ag,bennemanng} we see
that $\tau _{12}$ plays the role of the scattering time on magnetic
impurities. Thus, the intravalley superconductivity is completely destroyed
as soon as the inverse scattering intervalley time $\tau _{12}^{-1}$ becomes
larger than the transition temperature $T_{c}^{intra}$ in the absence of the
disorder.

It follows from the results obtained in the present and previous sections
that, by studying the possibility of superconductivity in graphene, it is
sufficient to concentrate on the intervalley pairing.

\section{Estimates}

Having derived the analytical expressions for the critical temperature
[Eqs.~(\ref{b26}) and (\ref{Tc})], we should determine now the parameters $\lambda $
and $\mu ^{\ast }$. Since we want to describe a graphene sheet where the
Fermi level can be tuned, we examine the dependence of $T_{c}$ on the
chemical potential $\mu $.

\begin{figure}[tbp]
\includegraphics[scale=0.6]{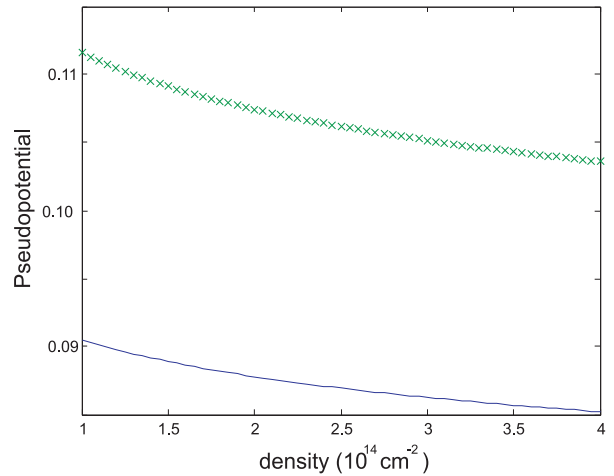}
\caption{(Color online) Pseudopotentials $\protect\mu_{11}$~(blue line) and $%
\protect\mu_{12}$~(green crosses) as function of the charge carrier density $%
n$.}
\label{mu}
\end{figure}

In order to estimate the Coulomb repulsion parameters $\mu _{11}^{\ast }$
and $\mu _{12}^{\ast }$ [Eqs.~(\ref{b25}) and (\ref{pseudopotential})], we use the
screened Coulomb potential [Eq.~(\ref{a7a})] and average this expression
over the Fermi surface in order to obtain $\nu V_{11}$ and $\nu V_{12}$.
These quantities allow us to calculate the Coulomb parameters $\mu
_{11}^{\ast }$ and $\mu _{12}^{\ast }$ as functions of the chemical
potential $\mu $ using Eqs.~(\ref{b25}) and (\ref{pseudopotential}). In order to
be specific, we have chosen $\kappa =2.5$ (Ref.~\onlinecite{Hwang}) for the value of the
effective dielectric permeability of the substrate (occupying halfspace)
entering Eq.~(\ref{a7}). This value corresponds to SiO$_{2}$. The dependence
of the parameters $\mu _{11}^{\ast }$ and $\mu _{12}^{\ast }$ on the
chemical potential $\mu $ (doping level) is represented for this value of $%
\kappa $ in Fig.~\ref{mu}. Further, we use $v_{0}=5.3\ \mathrm{eV}\mathring{%
\mathrm{A}}$ for calculations.

One can see from Fig. \ref{mu} that both pseudopotentials $\mu _{12}^{\ast }$
and $\mu _{11}^{\ast }$ decay slightly with increasing electron density $n$.

Calculation of the numerical values of the electron-phonon coupling constant
$\lambda $ is more difficult because one has to know exact values of the
matrix elements of the electron-phonon interaction. At the same time, the
electron-phonon coupling $\lambda $ determines the self-energy of the
electron-phonon interaction and can be extracted from photoemission studies.
Therefore, we simply take this value from literature.

For a long time, there has been a rather poor agreement between theoretical
results obtained using the local density approximation (LDA) and
experimental values concerning the total coupling strength and the ratio
between the two nonequivalent coupling parameters. The electron-phonon
coupling is also sensitive to the substrate (For details, see Ref.~[%
\onlinecite{Bianchi}] and citations therein).

A possible source of the disagreement has been identified in a recent paper,~%
\cite{siegel} where a copper substrate substantially screening the
electron-electron Coulomb interaction was used and the agreement between the
theoretical results~\cite{calandra} and the photoemission experiment was
found. Both the theory and the experiment with the copper substrate lead to quite low values of the electron-phonon coupling constant $\lambda $ that
remain below $0.05$ for electron densities up to $10^{14}\mathrm{cm}^{-2}$.
Using the metallic substrate one should assume that the dielectric
permeability of the substrate $\kappa $ entering Eq.~(\ref{a7}) is a
non-trivial function of the momentum. In order to avoid additional
calculations for this system we note that even setting $\mu _{12}^{\ast }=0$, 
the values $\lambda <0.05$ can not provide superconductivity with a
noticeable transition temperature.

Measurements of the electron-phonon coupling in potassium-doped graphene on
Ir$(111)$ substrate~\cite{Bianchi} have lead to the value $\lambda =0{.}28$
for a doping level of $\mu =1.29\ \mathrm{eV}$ (corresponding to the
electron density $n\approx 1\times 10^{14}\mathrm{cm}^{-2})$. Such a value
of $\lambda $ would lead to a rather high transition temperature. However,
the authors of Ref.~[\onlinecite{siegel}] argue that the assumption of the
linear spectrum used in Ref.~[\onlinecite{Bianchi}] leads to a considerably
overestimation of $\lambda $ and expect lower values of this parameter
corresponding to theoretical values of Ref.~[\onlinecite{calandra}].

The authors of Ref.~[\onlinecite{calandra}] suggest the following formula
for the function $\lambda \left( \mu \right) $ describing the dependence of
the electron-phonon coupling on the chemical potential:
\begin{equation}
\lambda \left( \mu \right) =5.55C\sqrt{n}10^{-9}\mathrm{cm}  \label{d2}
\end{equation}%
where $n$ is the number of electrons per surface area depending on $\mu$ via
$\mu=\sqrt{\pi n}$ and $C=1$.

This formula gives for $n=1\times 10^{14}\mathrm{cm}^{-2}$ the value $%
\lambda =0.056$, which perfectly agrees with the experimental results of
Ref.~[\onlinecite{siegel}] for graphene on the metallic substrate.

However, angle-resolved photoemission spectroscopy (ARPES) experiments \cite%
{chesney1,bostwick,chesney2} performed on doped graphene grown epitaxially
on SiC lead to considerably higher values of $\lambda $. Larger values of
the coupling constants obtained for graphene on other substrates mean that
the unscreened Coulomb interaction renormalizes the electron-phonon
interaction enhancing the latter. This conclusion correlates with the
results of Ref.~[\onlinecite{basko}], where an enhancement of the
intervalley electron-electron coupling constant $\lambda _{12}$ was
predicted. So, we can try to use the values of the coupling constant $%
\lambda $ obtained for such a non-metallic substrate.

According to a detailed analysis presented in Ref.~\onlinecite{chesney2},
the value of the coupling constant is $3.5-5$ times larger than predicted
theoretically~\cite{calandra}, which apparently implies that the coefficient
$C$ in Eq.~(\ref{d2}) should take the values $C\sim 3.5-5$. The dielectric
permeability of SiC equals $\kappa \approx 3{.}8$.~\cite{park}

\begin{figure}[tbp]
\includegraphics[scale=0.55]{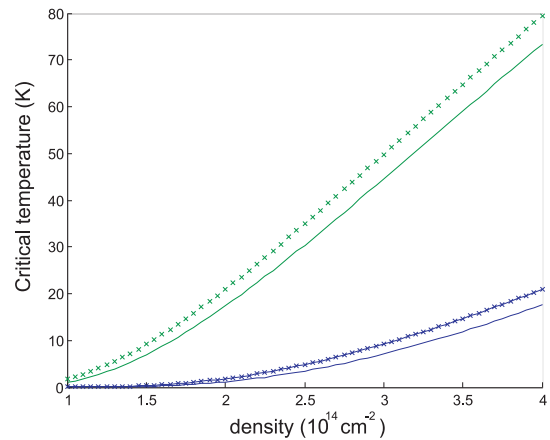}
\caption{(Color online) Critical temperature calculated with Eq.~(\protect
\ref{Tc}). The curves show $T_{c}$ as function of the electron density $n$.
Solid lines correspond to $\protect\kappa =2.5$ while dashed lines are curves
for $\protect\kappa =3.8$. The bottom blue lines correspond to $C=3.5$ and the top
green lines to $C=5.$}
\label{Tcfig}
\end{figure}

As the constants $C$ somewhat vary depending on the method of their
calculation and the pseudopotential $\mu _{12}^{\ast }$ depends on the
substrate, we simply draw in Fig.~\ref{Tcfig} the dependence of the critical
temperature $T_{c}^{inter}$ on the electron density $n$ for several values
of $C$ and $\kappa $ using Eqs.~(\ref{pseudopotential}, \ref{Tc}, \ref{d2}).
One can see from Fig.~\ref{Tcfig} that the superconductivity is possible for
realistic parameters characterizing the system and the transition
temperature $T_{c}^{inter}$ grows with increasing the electron density in
graphene. Using the maximal possible value for $C$, $T_{c}$ becomes very
high reaching the value of $70\ \mathrm{K}$ for very high electron
densities. This value of the critical temperature is apparently too high,
otherwise it would have been observed in the experiment.~\cite{efetov}
Therefore, the value $C=5$ does not look realistic. At the same time, the
value $C=3.5$ leads already to noticeable values of $T_{c}$.

The analysis presented above was done assuming that the chemical potential $%
\mu $ is far away from the Dirac points but is not close to the VHS.
According to Ref.~[\onlinecite{chesney1}], when approaching the VHS, the
electron-phonon coupling $\lambda $ grows very fast, which would further
increase the chances for the superconducting pairing. However, the
linear-band estimation method used in Ref.~[\onlinecite{chesney1}] was shown
to overstate the coupling,\cite{park} and the growth of the coupling $%
\lambda $ near VHS obtained in the latter publication was very slow reaching
the value $\lambda \approx 0.22$. This value would still be sufficient for
obtaining superconductivity with a reasonable critical temperature [see
Eqs.~(\ref{Tc})].

Unfortunately, the experiments \cite%
{Bianchi,siegel,chesney1,bostwick,chesney2} have not been supplemented by
transport experiments on the same materials at low temperatures, and it is
not clear whether the samples studied could be superconducting at low
temperatures, or not. At the same time, the superconductivity has not been
observed\cite{efetov} in the transport measurements on graphene with SiO$_{2}
$ substrate for the electron density $n$ up to $4\times 10^{14}\mathrm{cm}%
^{-2}$ and more efforts should be expended to clarify the situation.

\section{Discussion}

In this work, we estimated the superconducting transition temperature for
graphene as a function of the chemical potential $\mu $ or area electron
density $n$. We considered the case when the chemical potential is far away
from the Dirac point, which corresponds to very high electron density $n.$
At the same time, we assumed that the chemical potential $\mu $ is not in
the vicinity of the VHS.

Starting with a model describing the electron-phonon and Coulomb
interactions, we derived the Eliashberg equations for this system.
Considering both anomalous and normal self-energies, we have obtained
explicit formulas for the superconducting critical temperature that can be
used not only for a weak electron-phonon coupling $\lambda $, but also for $%
\lambda $ of order $1$. We show that the Coulomb interaction in graphene is
not very strong at high electron densities and does not necessarily destroy
the superconducting pairing.

As the parameters entering Eq.~(\ref{Tc}) are not precisely known, we have
drawn several curves in Fig.~\ref{Tcfig} corresponding to different values.
It is clear that the critical temperature rather weakly depends on the
dielectric permeability of the substrate and other details characterizing
the Coulomb interaction. At the same time, the dependence on the
electron-phonon coupling is strong, and we have shown curves corresponding to
different values of these constants that may be considered as realistic.

By estimating the pseudopotentials $\mu ^{\ast }$ describing the Coulomb
interaction and using values of the electron-phonon coupling $\lambda $
extracted from photoemission experiments, we come to the conclusion that the
transition temperature of the intervalley pairing $T_{c}^{inter}$ can reach
values exceeding $10\ \mathrm{K}$ for sufficiently high electron area
density $n$.

We have considered intra-and intervalley superconducting pairing and
demonstrated that the intervalley pairing is more favorable. Effect of
disorder on the intervalley superconductivity is weak but already a moderate
concentration of impurities can destroy the intravalley pairing. All this
means that the possibility of the intravalley pairing can be discarded in
realistic situations.

According to previous findings, the coupling constant $\lambda $ can be
considerably reduced provided a metallic substrate is used, which makes the
superconductivity improbable in such systems.

Intercalating graphene by various materials may lead to an additional source
of attraction between electrons and increase of the superconducting
transition temperature.

The superconductivity becomes even more favorable when approaching the VHS.
This is clear from the theoretical point of view because the density of
states diverges at this point, which should lead to a considerable increase
of $\lambda .$ The region of electron densities of order $10^{14}cm^{-2}$ is
apparently already rather close to the VHS. This would imply that the
approximation of the linear spectrum is no longer applicable. At the same
time, the Fermi velocity decreases when approaching the VHS, leading to an
additional increase of the density of states and, hence, of the critical
temperature $T_{c}$.

A slow growth of the electron-phonon coupling near the VHS obtained in Ref.~[%
\onlinecite{park}] using non-crossing self-energy diagrams indicates that
this divergency is missed in this calculation. Moreover, using only
non-crossing electron-phonon diagrams is not legitimate near the VHS and,
therefore, the analysis of Ref.~[\onlinecite{park}] is incomplete.

Provided the electron-phonon interaction grows and the interaction remains
essentially attractive one should expect at the VHS the conventional $s$%
-wave singlet superconductivity with a sufficiently high transition
temperature. By analyzing logarithmically diverging diagrams with the help of
renormalization-group equations, a new type of unconventional (chiral)
superconductivity was predicted recently\cite{chubukov,thomale} in the situation
when the interaction is repulsive. All this means that superconductivity in
graphene at high electron density is very probable and we hope that it will
be observed experimentally in the nearest future.

Strictly speaking, superconductivity with the identically zero resistance is
not possible in $2D$ due to fluctuations of the order parameter and a finite
energy required for generation of vortices. The transition temperature $%
T_{c} $ has been calculated in this work in the mean-field
approximation neglecting the fluctuations and vortices, and this is not
justified.

In practice, this means, however, that, instead of a sharp transition typical
for $3D$ superconductors, one would observe a slower decay of the
resistivity, which would make the transition rather broad. Although the
resistivity does not become exactly zero in such a superconducting state,
its value can be extremely small and not distinguishable from zero in real
experiments.

\section{Acknowledgements}

We thank D.K. Efetov for useful discussions and acknowledge a financial
support of Transregio 12 of DFG.


\begin{thebibliography}{99}
\bibitem{geim2004} K.S. Novoselov, A.K. Geim, S.V. Morozov, D. Jiang, Y.
Zhang, S.V. Dubonos, I.V. Grigorieva, and A.A. Firsov, Science \textbf{306},
666 (2004).

\bibitem{novoselov2005} K.S. Novoselov, A.K. Geim, S.V. Morozov, D. Jiang,
M.I. Katsnelson, I.V. Grigorieva, S.V. Dubonos, and A.A. Firsov, Nature
\textbf{438}, 197 (2005).

\bibitem{kim2005} Y. Zhang, Y.-W. Tan, H.L. Stormer, and P. Kim, Nature
\textbf{438}, 201 (2005).

\bibitem{castro} A.H. Castro Neto, F.\ Guinea, N.M.R. Peres, K.S. Novoselov,
and A.K. Geim, Rev. Mod. Phys. \textbf{81}, 109 (2009).

\bibitem{heersche} H.B. Heersche, P. Jarillo-Herrero, J.B. Oostinga, L.M.K.
Vandersypen, and A. Morpurgo, Nature \textbf{446}, 56 (2007).

\bibitem{uchoa} B. Uchoa, and A.H. Castro Neto, Phys. Rev. Lett. \textbf{98}%
, 146801 (2007).

\bibitem{kopnin} N.B. Kopnin and E.B. Sonin, Phys. Rev. Lett. \textbf{100},
246808 (2008); Phys. Rev. B \textbf{82}, 014516 (2010).

\bibitem{chesney} J.L. McChesney, A. Bostwick, T. Ohta, T. Seyller, K. Horn,
J. Gonzalez, and E. Rotenberg, Phys. Rev. Lett. \textbf{104}, 136803 (2010).

\bibitem{efetov} D.K. Efetov and P. Kim, Phys. Rev. Lett. \textbf{105},
256805 (2010).

\bibitem{gonzalez} J. Gonzalez, Phys. Rev B \textbf{78}, 205431 (2008).

\bibitem{chubukov} R. Nandkishore, L. Levitov, and A. Chubukov, arXiv:
1107.1903.

\bibitem{thomale} Maximilian Kiesel, Christian Platt, Werner Hanke,
Dmitry A. Abanin, Ronny Thomale, arXiv:1109.2953.

\bibitem{lozovik} Yu. Lozovik and A. Sokolik, Phys. Lett. A \textbf{374},
2785 (2010).

\bibitem{graphane} G. Savini, A.C. Ferrari, and F. Giustino, Phys. Rev.
Lett. \textbf{105}, 037002 (2010)

\bibitem{eliashberg} G.M. Eliashberg, Zh. Eksp. Teor. Fiz. \textbf{38}, 966
(1960); \textbf{39}, 1437 (1960) [Sov. Phys. JETP \textbf{11}, 696 (1960);
\textbf{12}, 1000 (1961)].

\bibitem{morel} P. Morel and P.W. Anderson, Phys. Rev. \textbf{125}, 1263
(1962).

\bibitem{mcmillan} W.L. McMillan, Phys. Rev. \textbf{167}, 331 (1968).

\bibitem{agd} A.A. Abrikosov, L.P. Gorkov, and I.E. Dzyaloshinskii, \textit{%
Methods of Quantum Field Theory in Statistical Physics, }Prentice Hall, New
York (1963).

\bibitem{grimvall} G. Grimvall, \textit{The Electron-Phonon Interaction in
Metals}, North Holland , Amsterdam (1980).

\bibitem{ferrari} A.C. Ferrari, J.C. Meyer, V. Scardaci, C. Casiraghi, M.
Lazzeri, F. Mauri, S. Piscanec, D. Jiang, K.S. Novoselov, S. Roth, and A.K.
Geim, Phys. Rev. Lett. \textbf{97}, 187401 (2006).

\bibitem{yan} J. Yan, Y. Zhang, P. Kim, and A. Pinczuk, Phys. Rev. Lett.
\textbf{98}, 166802 (2007).

\bibitem{basko} D.M. Basko and I.L. Aleiner, Phys. Rev. B \textbf{77},
041409 (2008).

\bibitem{Hwang} E.H. Hwang and S. Das Sarma, Phys. Rev. B \textbf{75},
205418 (2007).

\bibitem{vonsovsky} S.V. Vonsovsky, Yu. A. Izyumov, and E.Z. Kurmaev,
\textit{Superconductivity in Transition Metals, }Springer-Verlag, Berlin,
Heidelberg (1982).


\bibitem{carbotte} J.~P. Carbotte and F. Marsiglio, in \textit{The Physics
of Superconductors}, edited by K.~H. Bennemann and J.~B. Ketterson,
Springer-Verlag, Berlin (2003), p. 233.

\bibitem{aleiner} I. L. Aleiner, K. B. Efetov, Phys. Rev. Lett. \textbf{97},
236801 (2006).

\bibitem{bennemanng} L.~P. Gorkov, in \textit{The Physics of Superconductors}%
, edited by K.~H. Bennemann and J.~B. Ketterson, Springer-Verlag, Berlin
(2003), p. 347.

\bibitem{ag} A.A. Abrikosov and L.P. Gorkov, Zh. Eksp. Teor. Fiz. \textbf{39}%
, 1781 (1960) [Sov. Phys. JETP 12, 1243 (1961)]

\bibitem{Bianchi} M. Bianchi, E.D.L. Rienks, S. Lizzit, A. Baraldi, R.
Balog, L. Hornek\ae r, and Ph. Hofmann, Phys. Rev. B \textbf{81}, 041403(R),
(2010).

\bibitem{siegel} D.A. Siegel, C.G. Hwang, A.V. Fedorov, and A. Lanzara,
arXiv:1108.2566

\bibitem{calandra} M. Calandra and F. Mauri, Phys. Rev. B \textbf{76},
205411 (2007).

\bibitem{chesney1} J.L. McChesney, A. Bostwick, T. Ohta, K.V. Emtsev, T.
Seyller, K. Horn, and E. Rotenberg, arXiv:0705.3264

\bibitem{bostwick} A. Bostwick, T. Ohta, J.L. McChesney, T. Seyller, K.
Horn, and E. Rotenberg, Sol. St. Commun. \textbf{143}, 63 (2007).

\bibitem{chesney2} J.L. McCheseny, A. Bostwick, T. Ohta, K.V. Emtsev, T.
Seyller, K. Horn, and E. Rotenberg, arXiv:0809.4046

\bibitem{park} C.-H. Park, F. Giustino, J.L. McChesney, A. Bostwick, T.Ohta,
E. Rotenberg, M.L. Cohen, and S.G. Louie, Phys. Rev. B \textbf{77}, 113410
(2008).



\end{thebibliography}
\end{document}